\documentclass[epj]{svjour}
\usepackage{latexsym}
\usepackage{graphics}
\usepackage{url}
\begin{document}
\title{Signature of a possible $\alpha$-cluster state in $N=Z$ doubly-magic $^{56}$Ni}

\author{S.~Bagchi\inst{1,2,}\thanks{\emph{Corresponding author:} s.bagchi@gsi.de \newline \emph{Present address:} Indian Institute of Technology (Indian School of Mines), Dhanbad, Jharkhand - 826004, India} \and H.~Akimune\inst{3} \and J.~Gibelin\inst{4} \and M.~N.~Harakeh\inst{1} \and N.~Kalantar-Nayestanaki\inst{1} \and N.~L.~Achouri\inst{4} \and B.~Bastin\inst{5} \and K.~Boretzky\inst{2} \and H.~Bouzomita\inst{5} \and M.~Caama\~no\inst{6} \and L.~C\`aceres\inst{5} \and S.~Damoy\inst{5} \and F.~Delaunay\inst{4} \and B.~Fern\'andez-Dom\'inguez\inst{6} \and M.~Fujiwara\inst{7} \and U.~Garg\inst{8} \and G.~F.~Grinyer\inst{5} \and O.~Kamalou\inst{5} \and E.~Khan\inst{9} \and A.~Krasznahorkay\inst{10} \and G.~Lhoutellier\inst{4} \and J.~F.~Libin\inst{5} \and S.~Lukyanov\inst{11} \and K.~Mazurek\inst{12} \and M.~A.~Najafi\inst{1} \and J.~Pancin\inst{5} \and Y.~Penionzkhevich\inst{11} \and L.~Perrot\inst{9} \and R.~Raabe\inst{13} \and C.~Rigollet\inst{1} \and T.~Roger\inst{5} \and S.~Sambi\inst{13} \and H.~Savajols\inst{5} \and  M.~Senoville\inst{4} \and C.~Stodel\inst{5} \and L.~Suen\inst{5} \and J.~C.~Thomas\inst{5} \and M.~Vandebrouck\inst{14,15} \and J.~Van~de~Walle\inst{1,16}
%
}                     

%
%

\institute{KVI-CART, University of Groningen, NL-9747 AA, Groningen, The Netherlands \and GSI Helmholtzzentrum f\"ur Schwerionenforschung GmbH, 64291 Darmstadt, Germany \and Department of Physics, Konan University, Kobe 658-8501, Japan \and LPC Caen, ENSICAEN, Universit\'e de Caen, CNRS/IN2P3, Caen, France \and GANIL, CEA/DRF-CNRS/IN2P3, Bvd Henri Becquerel, F-14076 Caen, France \and Universidade de Santiago de Compostela, E-15706 Santiago de Compostela, Spain \and Research Center for Nuclear Physics, Osaka University, Osaka 567-0047, Japan \and Physics Department, University of Notre Dame, Notre Dame, Indiana 46556, USA \and IJCLab, Universit\'e Paris-Saclay, CNRS/IN2P3, 91405 Orsay Cedex, France \and Institute of Nuclear Research (ATOMKI), Debrecen, P.O.~Box 51, H-4001, Hungary \and Joint Institute for Nuclear Research, Dubna, Russia \and Institute of Nuclear Physics PAN, ul. Radzikowskiego 152, 31-342 Krak\'ow, Poland \and Instituut voor Kern- en Stralingsfysica, KU Leuven, B-3001 Leuven, Belgium \and IPN Orsay, Universit\'e Paris Sud, IN2P3-CNRS, F-91406 Orsay Cedex, France \and Irfu, CEA, Universit\'e Paris-Saclay, F-91191 Gif-sur-Yvette, France \and SCK CEN, Boeretang 200, 2400 Mol, Belgium}
\date{Received: date / Revised version: date}
%
\abstract{
An inelastic $\alpha$-scattering experiment on the unstable $N=Z$, doubly-magic $^{56}$Ni nucleus was performed in inverse kinematics at an incident energy of 50~A.MeV at GANIL. High multiplicity for $\alpha$-particle emission was observed  within the limited  phase-space of the experimental setup. This observation cannot be explained by means of the statistical-decay model. The ideal classical gas model at $kT$ = 0.4~MeV reproduces fairly well the experimental momentum distribution and the observed multiplicity of $\alpha$ particles corresponds to an excitation energy around 96~MeV. The method of distributed $m\alpha$-decay ensembles is in agreement with the experimental results if we assume that the $\alpha$-gas state in $^{56}$Ni exists at around $113^{+15}_{-17}$ MeV. These results suggest that there may exist an exotic state consisting of many $\alpha$ particles at the excitation energy of $113^{+15}_{-17}$ MeV.
%
} 
\authorrunning {S. Bagchi et al.,} 
\titlerunning {Signature of a possible $\alpha$-cluster state in $N=Z$ doubly-magic $^{56}$Ni}

\maketitle
\section{Introduction}
\label{intro}

The study of $\alpha$ clusters in nuclei is a very interesting field of research in nuclear physics. The main highlight of the $\alpha$-cluster studies is the Hoyle state (0$^{+}_{2}$) in $^{12}$C at 7.65 MeV. This resonance state, above the 3$\alpha$ threshold, is the doorway for the abundance of $^{12}$C in the universe through the nucleosynthesis process~\cite{Hoyle_Astro}, which resulted with life on earth.  In the 1960s, Ikeda et al.~\cite{Ikeda1968} predicted that $\alpha$-cluster states should exist closely above the particle-emission thresholds in light nuclei. In finite nuclei, the $\alpha$-cluster state may exist in excited states with a dilute density composed of a weakly interacting gas of $\alpha$ particles~\cite{Yamada2004,Itagaki2008}, where the degrees of freedom of the individual nucleons are no longer crucial~\cite{Freer_Nature}. Cluster structures in nuclei have been discussed in a wide range of theoretical models, such as, resonating group methods (RGM)~\cite{Wheeler}, generator coordinate methods (GCM)~\cite{Horiuchi}, self-consistent mean-field theory framework~\cite{Girod,Ebran}, \textit{ab-initio} calculations~\cite{Elhatisari}, Tohsaki-Horiuchi-Schuck-Röpke wave function~\cite{Tohsaki2001} etc. In a cluster state, the $\alpha$-clusters condense into the lowest S-wave orbit~\cite{Tohsaki2001} and thus, it is analogous to the Bose-Einstein condensed states of bosonic atoms. The condensation could occur in a low-density region. However, almost no experimental information on such states has been obtained except for light nuclei. It is now established that the lightest $\alpha$-conjugate system $^8$Be (it has been known for a long time that the ground state of this nucleus decays into two $\alpha$ particles) has a dumbbell configuration comprising two $\alpha$ particles~\cite{Arickx} separated by a distance of 4.4 fm~\cite{Tohsaki2017,Freer2018}. The search for the $\alpha$-cluster state was then extended towards the $\alpha$-conjugate nucleus $^{12}$C~\cite{Itoh2011} to enrich our understanding of the characteristics of the Hoyle-state~\cite{Hoyle_Astro,Smith,Aquila,Chernykh,Uegaki,Epelbaum}. In the case of $^{16}$O, the $\alpha$-cluster structure has been also studied~\cite{Wakasa2002,Wakasa2007} to explore experimentally the state in $^{16}$O equivalent to the Hoyle-state in $^{12}$C, which renewed the interest in this field~\cite{Funaki}. Recently, the $\alpha$-cluster state in $^{40}$Ca has been investigated~\cite{Borderie2016}. More details about the theoretical and experimental efforts to search and understand the cluster structures in nuclei have been elaborated in Refs.~\cite{Beck1,Beck2,Beck3}. 

The search for cluster structures in unstable nuclei became relevant in the past few decades with the availability of the radioactive ion-beam facilities and the state-of-the-art particle detectors. Inelastic $\alpha$-particle scattering has been found suitable to search for the condensate state because it has a selectivity to excite the natural parity states with $T$ = 0 in self-conjugate nuclei. Moreover, inelastic $\alpha$-particle scattering is a surface reaction, and the angular distributions are characteristic of the transferred angular momentum $\Delta L$. It has been successfully established in finding the condensate state in light nuclei~\cite{Itoh2011,Wakasa2007,Itoh_NPA,Freer2012}. The effort to study the $\alpha$-cluster state in unstable nuclei with inelastic $\alpha$-particle scattering has been reported in Refs.~\cite{Furuno,Akimune2013}.   

We present here the experimental results of inelastic $\alpha$-particle scattering off $^{56}$Ni, which suggest the signature of a possible $\alpha$ condensate in the unstable $N = Z$ doubly-magic $^{56}$Ni nucleus, potentially composed of 14$\alpha$ clusters. A preliminary report has appeared in Ref.~\cite{Akimune2013}.

\section{Experimental setup}
\label{setup}

We performed the experiment of inelastic $\alpha$-particle scattering off $^{56}$Ni ($T_{1/2}$ = 6.075 days~\cite{NNDC}) in inverse kinematics at GANIL. The $^{56}$Ni beam at 50 A.MeV incident energy was obtained via the process of nuclear fragmentation of a 75 A.MeV $^{58}$Ni beam bombarding a 526 $\mu$m thick Be target. The $^{56}$Ni beam was selected and purified using the ``Ligne d'Ions Super Epluch\'es" LISE spectrometer which consists of two dipole magnets with a 500 $\mu$m thick achromatic degrader (made of beryllium) \cite{Anne1987}. The secondary $^{56}$Ni beam intensity was $2\times10^4$ pps on average with a purity of $94.7\%$. The dominant contaminants were $^{55}$Co (purity $5.0\%$) and $^{53}$Fe (purity $0.3\%$). They were identified along with the ion of interest using the energy loss ($\Delta E$) and the time-of-flight information for low-intensity beam. The $\Delta E$ signal was derived from a Si detector placed before the MAYA active target, whereas the time-of-flight information was obtained using the trigger signals from the cyclotron radio frequency and a plastic scintillator placed before the MAYA setup. The Si detector, before the MAYA setup, was retracted from the beam line during the data-taking time with the high-intensity beam. In the offline analysis, the $^{56}$Ni beam was selected by putting a proper selection window in the time-of-flight spectrum; for details, see Ref.~\cite{Bagchithesis}.

Since the measurement was performed in inverse kinematics, an active target, MAYA \cite{Demonchy2007}, was employed. MAYA, developed at  GANIL, is a time-charge projection chamber with an active volume of $28(l) \times 25(w) \times 20 (h)$ cm$^3$ filled with helium gas (acting as $\alpha$-particle target) at 500 mbar pressure (with 5\% of CF$_{4}$ as a quenching gas). The pressure was adjusted to cover as much as possible all the ranges of the recoiling particles necessary for the study of inelastic $\alpha$ scattering in the excitation-energy range of 0 to 35 MeV, i.e. including the regions of the isoscalar giant monopole, quadrupole and octupole resonances at angles from 3$^\circ$ up to 8$^\circ$ in the center-of-mass (c.m.) frame. 

\begin{figure}
\resizebox{0.48\textwidth}{!}{%
  \includegraphics{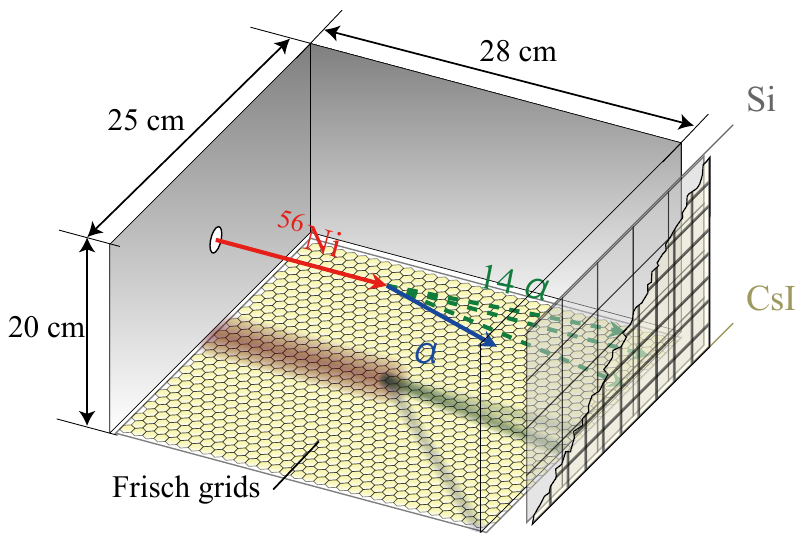}
}
\caption{(Color online) Schematic view of the experimental setup with the MAYA detector. A $^{56}$Ni beam (red) enters the active volume of MAYA and interacts with the helium gas resulting in the excitation of $^{56}$Ni and a recoil (scattered) $\alpha^\prime$ particle (blue). The $\alpha$ particles from the decay of $^{56}$Ni (dashed green) are forward focused and stop in the Si--CsI telescope. For readability, the mask below the beam path is not drawn here.}
\label{fig:setup}
\end{figure}

As shown in Fig.~\ref{fig:setup}, both the incoming $^{56}$Ni beam and the scattered $\alpha$ particle ionize the gas along their trajectories inside MAYA. The cathode plate at the top of MAYA was set to $-$3 kV, whereas the Frisch grid was maintained at ground potential. Due to the applied electric field between the Frisch grid and the cathode plate, electrons, generated in the ionization process, drift towards the Frisch grid. Furthermore, a high voltage of +1.3 kV was applied to 32 amplification wires that were below the Frisch grid and were parallel to the beam direction. The avalanches on the wires induced signals on a matrix of $32 \times 32$ hexagonal pads of 5 mm side length, which were individually recorded \cite{Demonchy2007}. Thus, a two-dimensional picture of the reaction was reconstructed on the pad plane. The drift velocity of electrons was 1.7 cm/$\mu$s. The drift time of electrons to each wire was recorded in order to determine the reaction plane in the three-dimensional space. 

\begin{figure}
\resizebox{0.48\textwidth}{!}{%
  \includegraphics{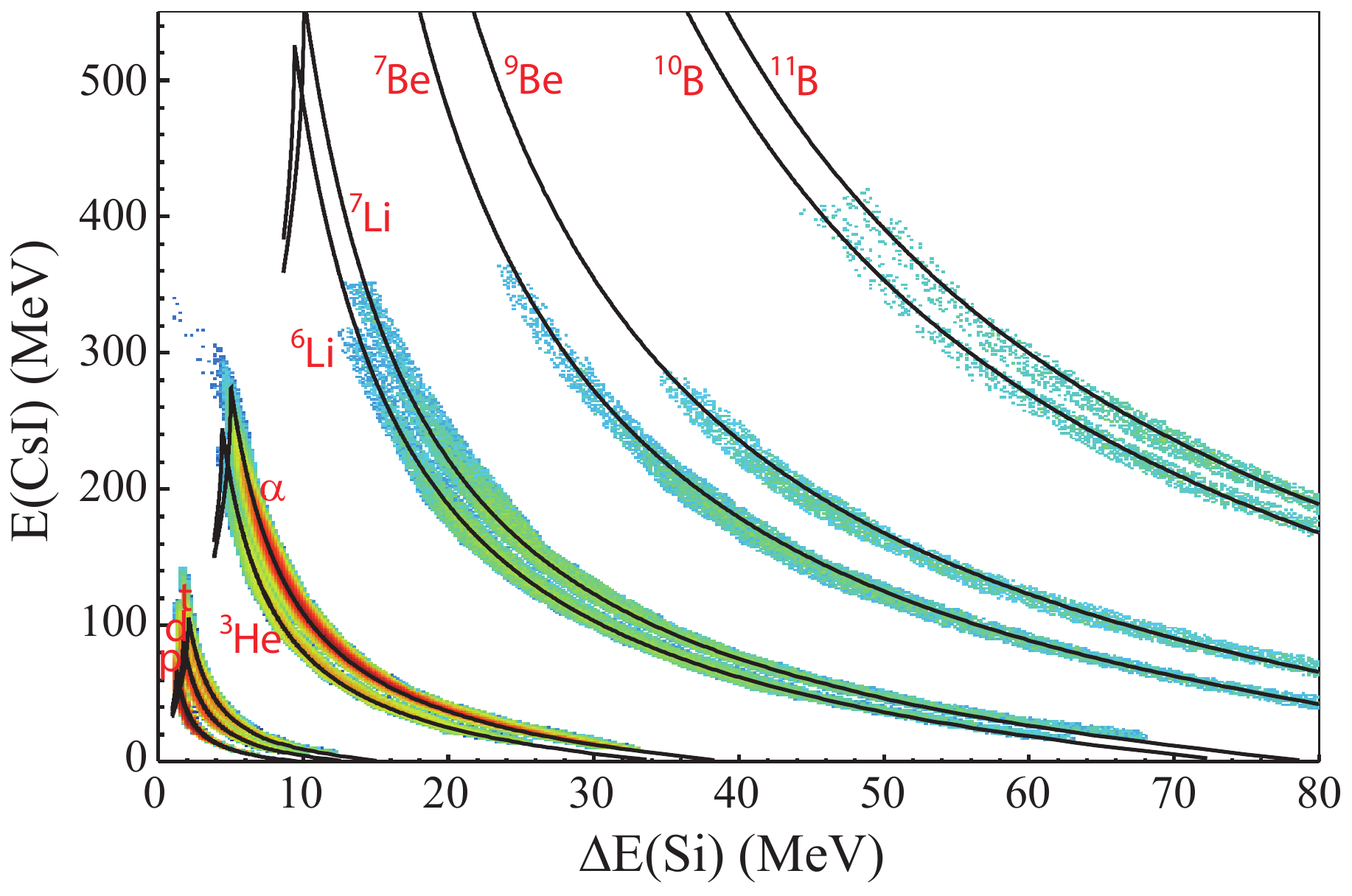}
}
\caption{(Color online) Two-dimensional scatter plot of energies deposited by charged particles in the CsI detectors versus energies deposited in the Si detectors after CsI light-output correction. The experimental loci are compared with theoretical calculations for light particles.}
\label{fig:EdE}
\end{figure}

In order to measure both the recoil particle and the beam efficiently, an electrostatic mask~\cite{Pancin2012} was placed 1~cm below the beam axis in MAYA (not shown in Fig. 1) to reduce the number of electrons produced through the high ionization by the $^{56}$Ni beam and thereby reduce space-charge effects. The mask was extended through the central region of the cathode pad plane along the beam direction and did not interfere with the electric field where the recoil $\alpha$ particles were detected. It consisted of 9 wires of 50 $\mu$m diameter with a 1 mm pitch. The three central wires were biased at $-$900 V, the next two wires on each side were biased at $-$1000 V, and the outermost wires were biased at $-$1100 V (See Fig. 3.13 of Ref.~\cite{Bagchithesis}) to capture electrons originating from the beam ionization. The outer-most wires were kept at the same potential as that generated by the applied drift electric field at the height of the wire mask. Therefore, the amount of charge collected on the central wire due to the ionization by the beam is greatly reduced and the detection of charge produced by the less ionizing recoil $\alpha$ particles becomes more sensitive.

As shown schematically in Fig.~\ref{fig:setup}, 20 $\Delta E$-$E$ telescopes, each consisting of one 700~$\mu$m thick Si detector having dimensions of $5 \times 5$~cm$^2$ ($\Delta E$) followed by four 2-cm thick CsI scintillators ($E$) with surface area of 2.5 $\times$ 2.5 cm$^2$ (resulting in 80 separate $\Delta E$-$E$ segments), were placed 5~cm downstream of the active region. The 20 $\Delta E$-$E$ telescopes were arranged in a 5 (horizontal) $\times$ 4 (vertical) matrix, and allowed the identification of all light charged particles with kinetic energies high enough to pass through the Si layer. The pre-amplifiers of the Si detectors were connected to CAEN N568B spectroscopy amplifiers with dual outputs comprising a normal signal and a signal amplified by a factor of ten, which allowed us to achieve a better resolution at low energy (typically for $p$, $d$, and $t$ identification). Because we had only one Si detector followed by four CsI detectors, we deduced the corresponding $\Delta E$ energy loss in the Si detector by dividing the measured total energy with the number of CsI-detector hits at the back. This approach was validated with simulations. Energy calibration for the Si detectors was performed up to 7.7~MeV using a $^{226}$Ra 4$\alpha$-line source. The responses of the CsI detectors were calibrated from the extrapolation of the energy deposited in the Si detectors. The light-output correction for the CsI detectors was performed following the procedure described in Ref.~\cite{Parlog2002}. The resulting two-dimensional scatter plot of deposited $E$ versus $\Delta E$ energies is presented in Fig.~\ref{fig:EdE}. The total energy deposited by an $\alpha$ particle is then obtained as the sum of the $\Delta E$ and $E$ signals.

One must note that the aim of this experiment was originally to search for the compression modes in $^{56}$Ni using inelastic $\alpha$-particle scattering in inverse kinematics~\cite{Bagchithesis,Bagchi2015}. The detailed discussions about the setup, the calibration and analysis procedures are given in Refs.~\cite{Bagchithesis,Roger2011}.

\section{Analysis}
\label{Analysis}

\begin{figure}
\resizebox{0.48\textwidth}{!}{%
  \includegraphics{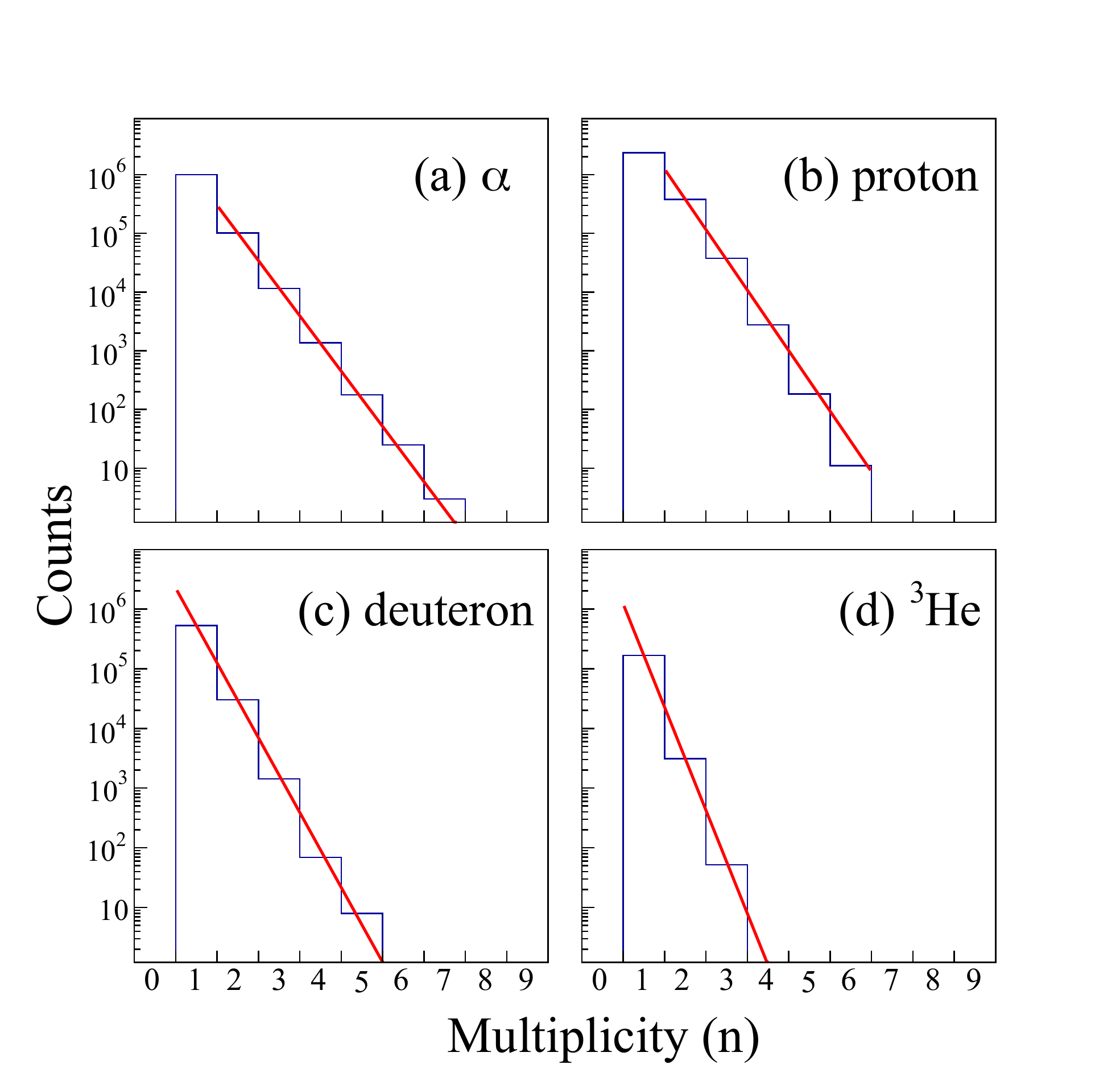}
}
\caption{(Color online) Multiplicity distributions of $\alpha$ particle (a), proton (b), deuteron (c) and $^3$He (d). The histograms and solid lines correspond to the experimental data and the results of fitting, respectively.}
\label{fig:mult:all}
\end{figure}

Due to the limitation of the phase-space covered by the experimental setup, excitation energies of $^{56}$Ni only up to $\sim$ 35 MeV could be reconstructed using the missing-mass method. Hence, the extraction of the resonance energy of a possible $\alpha$-condensate state in $^{56}$Ni, which is well above 35 MeV excitation energy, is not feasible experimentally in the present setup. The analysis of the particle-emission data was performed here with the wall of solid-state detectors as shown in Fig.~\ref{fig:setup}. In this analysis, we found that many particles are emitted at forward angles with high multiplicities. In Fig.~\ref{fig:mult:all}, the distributions of the observed multiplicities ($n$) of $\alpha$ particle, proton, deuteron, and $^{3}$He are shown. The pile-up events due to multiple particles hitting the same CsI crystal were simply discarded. Note that experimentally the number of pile-up events was negligible because of the limited intensity of the incoming $^{56}$Ni beam. This can also be deduced from Fig.~\ref{fig:EdE}, which shows no evidence of events between the loci of the different ejectile nuclei and thus confirming the absence of pile-up events in case of the present analysis. As can be seen in Fig.~\ref{fig:mult:all}, the maximum multiplicity observed for $\alpha$ particles is seven. This means that in the limited phase-space acceptance of the present experiment, half of the mass of $^{56}$Ni is observed to be carried away by means of emitted $\alpha$ particles. The multiplicity distributions are well reproduced by an exponential function $\exp\left(-\lambda n\right)$, where $n$ is the measured multiplicity. The multiplicity slope $\lambda$ is determined to be $2.17\pm0.07$ for $\alpha$ particles.

The $\lambda$ value for $\alpha$-particle emission is the smallest compared to the multiplicity slopes for proton, deuteron, and $^3$He, which are $2.37\pm0.11$,  $2.88\pm0.21$, and $3.98\pm0.32$, respectively. The parameter $\lambda$ reflects the dependence of the measured $\alpha$ multiplicity, $n$, on the cross section for a total decay of $m\alpha$ particles in the $^{56}$Ni($\alpha,\alpha^{\prime} m\alpha$) reaction and on the experimental acceptance. This experimental result is surprising, because it is natural to expect that the charged-particle decay from states at high excitation energies is suppressed by the Coulomb barrier relative to neutron decay. Furthermore, one would expect $\alpha$ emission to be more suppressed than proton and deuteron emission. In fact, the maximum multiplicities of protons and deuterons are smaller than that of $\alpha$ particles. The observation of multiple $\alpha$-particle decay in the present experiment would be a signature of a possible existence of an $\alpha$-condensed state in $^{56}$Ni at high excitation energy.

\section{Statistical-decay model}
\label{statistical}

In order to examine the possibility of statistical decay from compound states, which may have a small overlap with an $\alpha$-cluster state, we performed statistical-decay model calculations based on the Hauser-Feshbach formalism using the code CASCADE \cite{Puhlhofer1977}, which was modified to take spin, isospin and parity exactly into account in the calculations \cite{Harakeh1986}. In that respect, global level-density parameters were used to describe the unknown levels at high excitation energies in daughter nuclei applying proper spin, isospin, and parity quantum numbers to these input parameters. Furthermore, individual level parameters for the low-excitation-energy region were taken from the experimental data. All the decay channels, i.e. ${\gamma}$, neutron, proton, and $\alpha$ decays, are taken into account applying the proper spin and isospin Clebsch-Gordan coefficients coupling decaying states to final states. The spin, parity and isospin for the initial decaying states in $^{56}$Ni that have been excited by inelastic $\alpha$ scattering were set to be $J^\pi=0^+$ and $T=0$, assuming that a monopole resonance is strongly excited at high excitation energies around 100 MeV (see the next section for detailed explanation). However, it should be noted that the performed statistical-model calculations show that results do not change significantly if we assume a low-spin initial state instead of a monopole state. 

\begin{figure}
\resizebox{0.48\textwidth}{!}{%
  \includegraphics{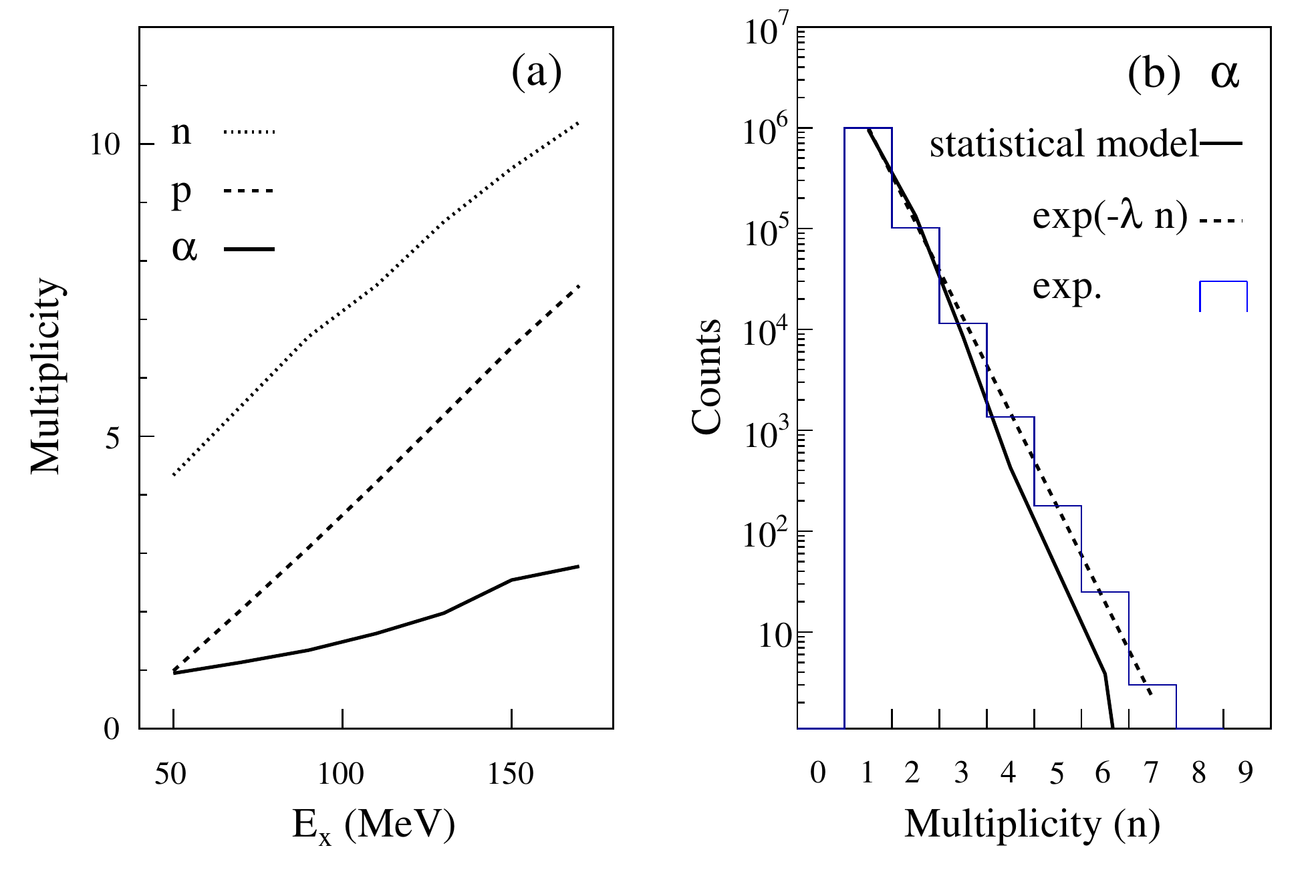}
}
\caption{(Color online) (a) Excitation-energy dependence of average multiplicities estimated using the CASCADE code for the emission of $\alpha$ particles (solid line), protons (dashed line), and neutrons (dotted line). (b) Multiplicity distribution of $\alpha$ particles at E$_x$ = 110 MeV: (solid line) estimated using the CASCADE code, (histogram) experimental data and (dashed line) result of fitting experimental data with an exponential function.}
\label{fig:casc:mul}
\end{figure}

The maximum angular momentum for decay products was set to be $40\hbar$. Figure~\ref{fig:casc:mul}(a) shows the average multiplicities of protons, neutrons, and $\alpha$ particles estimated in the statistical-decay model using the CASCADE code. Even for an excitation energy, $E_{x}$, of about 150~MeV, the estimated average multiplicity for decay $\alpha$ particles is found to be significantly less than five. In the statistical-model calculations, competition between different decay channels is automatically taken into account.  We have performed statistical-model calculations for decay  of $^{56}$Ni excited to 110 MeV (Fig.\ref{fig:casc:mul}(b)) using CASCADE taking spin, isospin and parity properly into the calculations. These calculations have been used to estimate the number of $\alpha$ particles emitted by statistical decay from the excited state of $^{56}$Ni. The CASCADE code is not a Monte-Carlo statistical code, and therefore the decay chains cannot be followed separately. We have deduced the $\alpha$ multiplicity from the population of $^{52}$Fe, $^{48}$Cr, $^{44}$Ti, $^{40}$Ca, $^{36}$Ar, $^{32}$S, and $^{28}$Si. In the statistical-model calculations with the CASCADE code, one cannot distinguish between an $\alpha$ emission or 2$p$-2$n$ emission through which a nucleus is populated, although the energetics are quite different, since 2$p$-2$n$ corresponds to full breakup of an $\alpha$ particle requiring 28.3~MeV extra energy. Due to these processes, the statistical model calculation overestimates the $\alpha$ multiplicities. This implies even a larger discrepancy with the experimental data.

The lightest nuclei produced by the decay were found to be $^{32}$S, $^{31}$S and $^{31}$P. Lighter nuclei than these are produced with a probability of less than 0.1$\times$10$^{-7}$, which is the minimum of the calculation accuracy of CASCADE. The probability of emitting six $\alpha$ particles sequentially was estimated to be 0.7$\times$10$^{-6}$. In Fig.~\ref{fig:casc:mul}(b), the total number of counts in the statistical model is normalized to the experimental observation at multiplicity 1. Therefore, with the above-mentioned probability in the statistical model, the total number of counts for multiplicity six is estimated to be around 0.7. On the other hand, the experimentally observed number of counts for multiplicity six is around 20, which is about a factor 30 higher than the statistical-model prediction. The difference is even higher in the case of multiplicity seven. In Fig.~\ref{fig:casc:mul}(b), there is a sharp kink in the statistical-model curve showing the total number of counts is decreasing rapidly to zero for multiplicities larger than five. Hence, the absolute discrepancy between the statistical model and the experimental observation is large as is also evident from the average multiplicity in Fig~\ref{fig:casc:mul}(a). Furthermore, in the sequential decay chain of $\alpha$ particles, a proton or a neutron can be emitted preceding or following one of the $\alpha$ emissions. The probability of emitting six $\alpha$ particles and one proton is estimated to be 0.4$\times$10$^{-6}$, and the probability of emitting six $\alpha$ particles and one neutron is estimated to be less than 0.1$\times$10$^{-7}$. In fact, these background processes add up to the pure sequential $\alpha$ decay and make the probability of six-$\alpha$ emission a bit larger in the statistical model. This, together with the overestimation of the $\alpha$ multiplicities due to the 2$p$-2$n$ emission discussed above, implies that the $\alpha$ multiplicities calculated in the statistical model with CASCADE should be (much) lower than shown in Fig.\ref{fig:casc:mul}(b) and therefore highly underestimate the observed $\alpha$ multiplicities.

The decay constant of the probability of emitting n $\alpha$ particles via the statistical decay processes were estimated to be 2.49. Therefore the statistical decay model cannot explain the value 2.17 $\pm$ 0.07 obtained in the experiment. Moreover, the statistical-model calculations are not able to explain the highest observed $\alpha$ multiplicities. 

\section{Ideal-gas model}
\label{idealgas}

One of the possibilities to explain the observation of such a high multiplicity of $\alpha$ particles is to introduce the concept that many-$\alpha$-particle emission may come from an exotic high-lying state such as a dilute $\alpha$-gas state. In order to understand qualitatively whether this hypothesis is correct, we performed a simulation based on the simple statistical model. In this model, we have assumed that the excited state in $^{56}$Ni consists of fourteen independent $\alpha$ clusters, and that the distribution of the kinetic energy for each $\alpha$ particle follows the Boltzmann distribution for the classical ideal gas at a relatively low temperature $T$. The highly excited $^{56}$Ni nuclei are assumed to move along the $z$ direction with  a kinetic energy of 50 A.MeV and emit fourteen $\alpha$ particles. The kinetic energy and the direction in the c.m.~system of the produced $^{56}$Ni nuclei were randomly given by assuming the Boltzmann distribution. We have used the Boltzmann distribution to describe the decay of the excited $\alpha$ condensate. In principle, the Bose-Einstein distribution should have been used. However, this is beyond the scope of this article for two reasons. The experimental reason has to do with our experimental detection system, which does not have the necessary granularity for high spatial resolution needed for forward-boosted $\alpha$ particles due to the high velocity of the c.m. system. This makes differentiation between models difficult. The second reason for adopting the classical ideal gas model is that classical statistics have only one free parameter ($kT$). Using the Bose-Einstein distribution instead requires fitting more parameters, which is not justified by the quality of our experimental data. This is an experimental task for the future. 

In the simulation, the detection probability was estimated event by event for emitted $\alpha$ particles. We reconstructed the deposited energy in the Si+CsI array by taking into account its segmentation. We also simulated the thickness of the $^{4}$He gas target (28 cm in length) by randomly choosing the vertex point of the reaction. The polar and azimuthal angles of the recoil $\alpha$ particles are then reconstructed accordingly.

\begin{figure}
\resizebox{0.48\textwidth}{!}{%
  \includegraphics{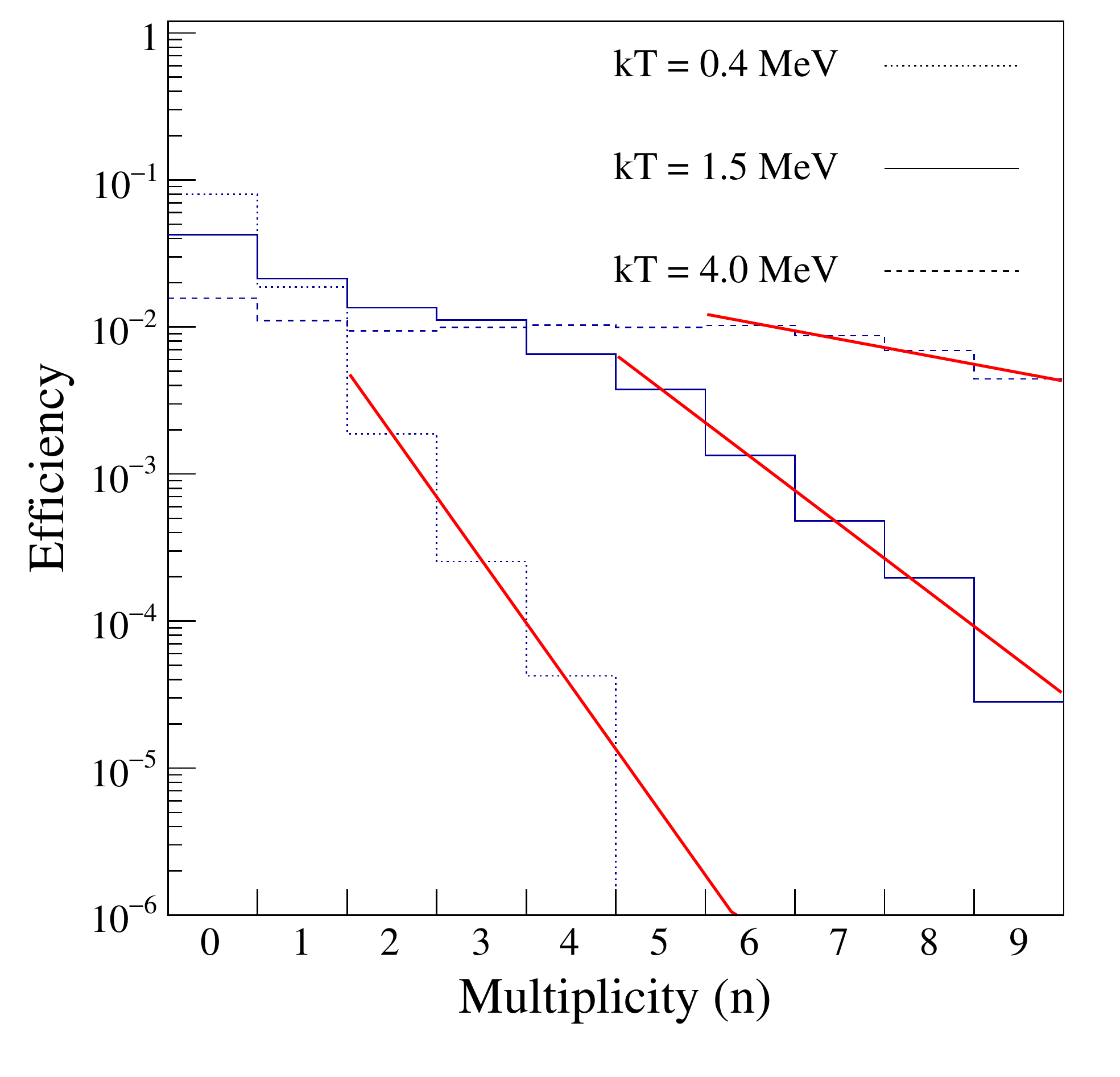}
}
\caption{(Color online) Estimated  detection efficiency for a fourteen-$\alpha$ event at $kT =0.4$ MeV (dotted  histogram  ), $1.5$ MeV (solid histogram), and $4.0$ MeV (dashed histogram). The solid lines are the results of fitting with an exponential function $\exp\left(-\lambda n\right)$.}
\label{fig:det:eff}
\end{figure}

Figure~\ref{fig:det:eff} shows the results of the simulation for the detection efficiency of $\alpha$ particles from $^{56}$Ni at $kT= 0.4, 1.5,$ and $4.0$~MeV, where $k$ is the Boltzmann constant. The variable $n$ is defined as the number of $\alpha$ particles detected within the limited phase-space acceptance of the present experimental setup. In the simulation, pile-up events were discarded as has been done in the experiment. When the simulation is carried out with the temperature of $kT$ = 0.4 MeV, the corresponding parameter $\lambda = 1.99\pm0.8$ is in close agreement with the experimentally determined value for $\alpha$ particle emission within the errors. The temperature of 0.4~MeV corresponds for $m=14$ to an excitation energy of $\frac{3}{2} mkT=8.4$~MeV above the fourteen-$\alpha$ separation threshold energy in $^{56}$Ni. Since the 14$\alpha$ separation threshold energy is $E_{th}^{14} = 87.9$~MeV, this means that the condensate is close to it at a corresponding excitation energy of $E_{x}\approx 96 $~MeV.

In Fig.~\ref{fig:det:eff}, we see that the efficiency for detecting $\alpha$ particles becomes lower with the increasing number of detected $\alpha$ particles. Furthermore, the efficiency is especially low at lower temperatures, because at the lower temperatures the $\alpha$ particles move with small kinetic energies in the c.m.~system of the highly excited $^{56}$Ni nucleus, and are therefore boosted to very small angles in the laboratory system. Since no particle detectors were placed on the incident-beam axis, it was impossible for us to measure $\alpha$ particles scattered in the very forward direction close to the beam axis. Also, when more than one $\alpha$ particle (pile-up event) enter the same detector telescope, it is excluded from the number of $\alpha$ particles detected in the simulation since it cannot be identified.

Since the spread of the $\alpha$ particles in the $x-y$ direction in the laboratory system becomes small when the temperature is low ($kT < 1$~MeV), the detection efficiency sharply decreases with increasing the number of detected $\alpha$ particles as shown in Fig.~\ref{fig:det:eff}. At $kT = 0.4$~MeV, the detection efficiency for an event with seven $\alpha$ particles detected out of fourteen in total is $<10^{-4}$\%. Taking this into account, the cross section for excitation of the $\alpha$-condensate state via inelastic $\alpha$ scattering is estimated to be unexpectedly large for $kT = 0.4$ MeV, although the experimental multiplicity slope $\lambda$ is well reproduced.

At $kT =1.5$ and $4$~MeV, the detection efficiency for an event with seven $\alpha$ particles detected out of fourteen in total are 0.06\% and 0.8\%, respectively. At high $kT$, the detection efficiency cannot be reproduced by a simple exponential function and the probability of detecting $\alpha$ particles of 4 or less does not depend strongly on the number of detected $\alpha$ particles. This is because the angles in the laboratory system of the emitted $\alpha$ particles become on the average larger for lower multiplicities, so that more $\alpha$ particles are detected in the counter telescopes.

\section{Distributed $m \alpha$-decay ensemble model}
\label{malpha}

\begin{figure}
\resizebox{0.48\textwidth}{!}{%
  \includegraphics{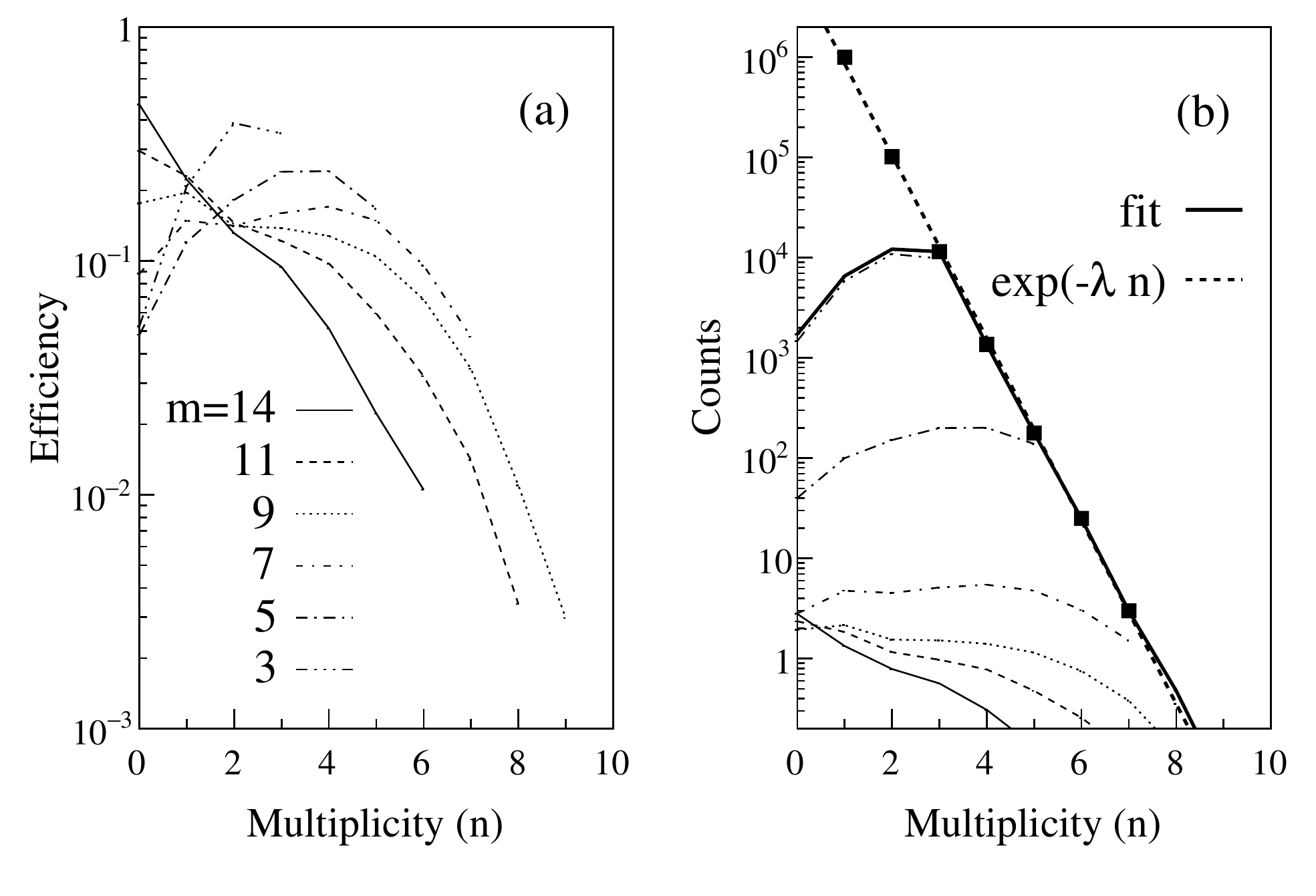}
}
\caption{(a) Detection efficiencies for $m$ = 14 (solid), 11 (dashed), 9 (dotted), 7 (dot-dashed), 5 (dot-long-dashed), and 3 (dot-dot-dot-dashed) $\alpha$ particles are shown. See text for definition. (b) Experimentally observed multiplicities for $\alpha$ particles are shown in black filled squares. Fit to the experimental data with efficiencies for $m$ = 14 to 3 $\alpha$ particles is shown with the thick solid line. Decomposition of partial contributions of $m$ = 14 (solid), 11 (dashed), 9 (dotted), 7 (dot-dashed), 5 (dot-long-dashed), and 3 (dot-dot-dot-dashed) $\alpha$ particles is shown. For comparison, exponential fitting ($e^{-\lambda n}$) with $\lambda = 2.17\pm0.07$ to the data is also shown (thick dashed). }
\label{fig:det:fit}
\end{figure}

In the de-excitation process of the $\alpha$-cluster state in $^{56}$Ni, it is natural to assume that there is a process for decay not only by emitting 14$\alpha$ particles but also by emitting multiple ($<14$)$\alpha$ particles decaying to residual nuclei. We investigated whether the process of decaying into a number of $\alpha$ particles can be understood by using a model that considers such a decay process, i.e. a decay process in which $m\alpha$ particles are emitted from an excited state in $^{56}$Ni leading to a residual nucleus with a mass number $A = 4(14 - m)$. Therefore, according to this scenario, we assume that the multiplicity distribution of $\alpha$ particles observed in the experiment is due to an ensemble of various $m\alpha$ decay processes. The excitation energy of $^{56}$Ni was obtained from the theoretical value calculated by Yamada and Schuck \cite{Yamada2004}  based on the dilute $m\alpha$-condensate model. They estimated the excitation energy of the $m\alpha$-cluster state up to $m = 12$. The excitation energy of the $\alpha$-cluster state for $m = 14$ system is simply estimated extrapolating their result to be $E_{T}^{14} = 25$ MeV above the $14\alpha$ separation threshold energy of $E_{th}^{14} = 87.9$ MeV.

\begin{figure}
\resizebox{0.48\textwidth}{!}{%
  \includegraphics{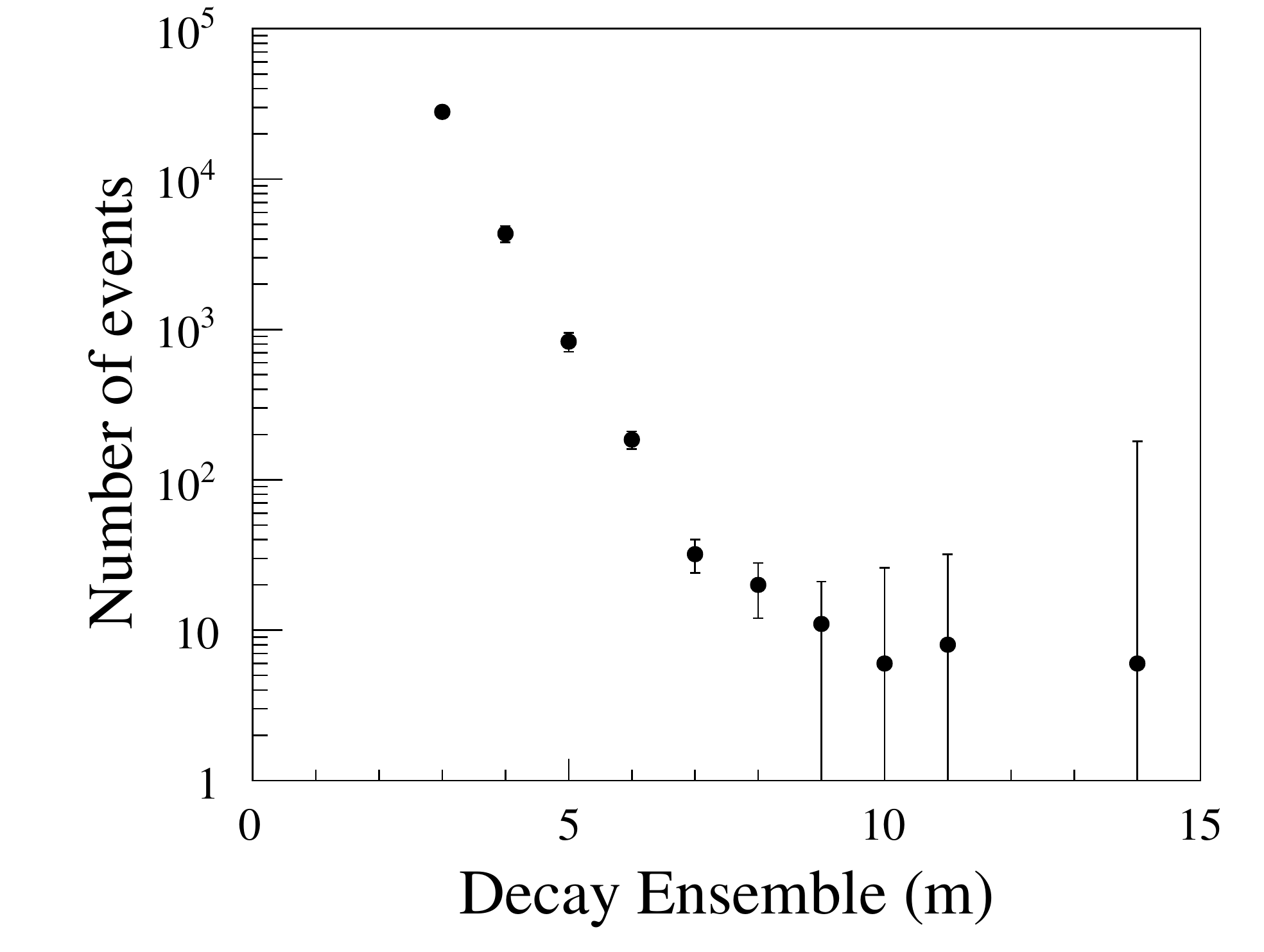}
}
\caption{Plot of the number of source events for the $m\alpha$-emission channel decomposed as shown in Fig.~\ref{fig:det:fit}(b).}
\label{fig:sim:sum}
\end{figure}

In the following discussion, the excitation energy $E_{x}$ is fixed to be $E_{x}=E_{th}^{14}+E_{T}^{14}=113$~MeV and it is assumed that $m\alpha$-particles share the energy $E_{T}^{m} = E_{x}-E_{th}^{m}$ according to the Boltzmann distribution. Here, $E_{T}^{m}$ is the temperature of $m\alpha$ particles and $E_{th}^{m}$ is the threshold energy for separation of $(m)\alpha$ particles resulting in a nucleus with mass number $A = 4(14 - m)$ as residue. The recoil energy in the c.m. system was not taken into account. This is a reasonable simplification at low multiplicities for which the recoil energy is small and will have little effect on the emitted $\alpha$ particles since the $\alpha$ particles are boosted with high energies in the forward direction. At high multiplicities, the $\alpha$-particle emission has no preferred direction and therefore the recoil energy of the residue is on the average small. When the number of emitted $\alpha$ particles is one or two, the decay process is not necessarily that of a state having an $\alpha$-cluster structure, but also includes the case of a statistical decay process. We did not take into account $\alpha$-particle emission processes for estimation of the $\alpha$-cluster structure. We define here $n$ as the number of $\alpha$ particles detected within the limited phase-space acceptance of the present experimental setup and $m$ as the number of $\alpha$ particles emitted from the excited states of $^{56}$Ni. Figure~\ref{fig:det:fit}(a) shows the results of simulations of detection efficiencies for $m =14, 11, 9, 7, 5, 3$ $\alpha$-particle decay events. Using these results, we determined the branching ratio for decay of $m\alpha$ particles using the formula,

\begin{equation}
Y_{n} = \sum_{m=3}^{14} X_{m} \epsilon_{m}(kT, n),
\end{equation}
to fit the experimental data. Here, $X_{m}$ is the estimated number of source events for $m\alpha$-particle decay, $\epsilon_{m}(kT, n)$ is the detection efficiency for $n\alpha$ particles out of $m\alpha$ particles at temperature $T$ in $m\alpha$-particle decay, and $Y_{n}$ is the yield of $\alpha$ particles for observed multiplicity $n$. The multiplicity distributions of $\alpha$ particles are well described with the sum of decay branches (see Fig.~\ref{fig:det:fit}(b)).

Figure~\ref{fig:sim:sum} shows the $X_{m}$ plot. Since events up to $n$~=~7 were measured in the experiment, the branching ratios up to $m$~= 7 are well determined for the given $E_{th}$. Furthermore, since no events with $n$~$>$ 7 were detected, the estimated number of source events for $m\alpha$-particle decay is not well determined. Although the simulation of the detection efficiency of multi-$\alpha$ particles varies depending on $E_{x}$ as described above, the experimental data are well fitted even when the excitation energy $E_{x}$ is varied between
8 and 40 MeV. Therefore, an uncertainty remains in determining the excitation energy in this experiment in the range of $E_{T} ^{14}$ = 8 to 40 MeV. Since the extrapolation gives $E_{T}^{14}$ to be 25~MeV, the associated errors can be assigned to the excitation energy, that is $113^{+15}_{-17}$~MeV.

\begin{figure}
\resizebox{0.48\textwidth}{!}{%
  \includegraphics{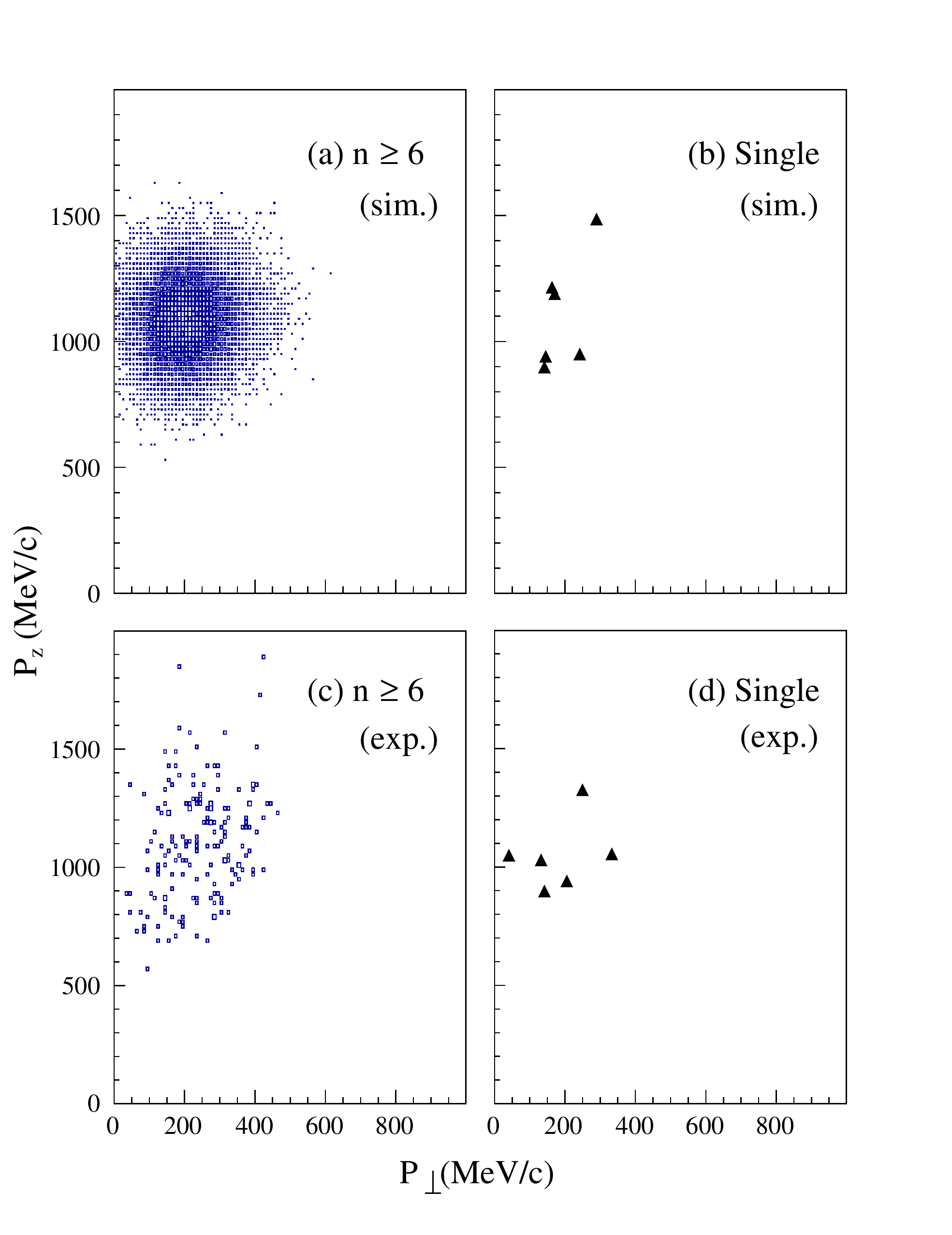}
}
\caption{(Color online) Two-dimensional scatter plots of the momentum distributions of $\alpha$ particles. Results of simulations based on the classical ideal-gas model from events with multiplicity $\geq 6$ (a), and from a single event with multiplicity six (b). Experimental momentum distributions of $\alpha$ particles from events with multiplicity $\geq 6$ (c), and from a single event with multiplicity six (d).}
\label{fig:simu_expt}
\end{figure}

We performed additional simulations in order to confirm the validity of the present approach. Figures~\ref{fig:simu_expt}(a) and (b) show the two-dimensional scatter plots of the momentum distributions of $\alpha$ particles obtained by simulations at $E_{T}=25$~MeV. Figure~\ref{fig:simu_expt}(a) shows the superposition of all the events with a multiplicity $\geq 6$.  Figure~\ref{fig:simu_expt}(b) shows an example of the momentum distributions for a single event with multiplicity six. Figures~\ref{fig:simu_expt}(c) and (d) are the same as Figures~\ref{fig:simu_expt}(a) and (b), respectively, but they are obtained in the experiment. The average momentum in the $z$ direction (perpendicular to the $z$ direction) obtained in the experiment for multiplicity $n\geq6$ is $\bar{p}_{z}=1100$~MeV/c ($\bar{p}_{\perp}=240$~MeV/c) and its standard deviation is $\sigma_{z}=230$~MeV/c ($\sigma_{\perp}=100$~MeV/c). In the simulation, the energies of the $\alpha$ particles are normalized to fit the experimental result. Therefore, we used for the simulations $\bar{p}_{z}=1100$~MeV/c, $\sigma_{z}=230$~MeV/c, $\bar{p}_{\perp}=240$~MeV/c and $\sigma_{\perp}=100$~MeV/c. It can be seen that after the normalization the simulation reproduces the momentum distributions obtained from the experiment rather well. 

\section{Summary and conclusion}
\label{summary}
We performed the $\alpha$($^{56}\rm{Ni},\alpha^{\prime}$) inelastic scattering experiment in inverse kinematics at an incident energy of 50 A.MeV. Multi $\alpha$-decay events were observed with a maximum multiplicity of seven particles within the limited phase-space acceptance of the present experimental setup. This result cannot be described by performing a simple calculation in the framework of the statistical-decay model and suggest a signature of the possible existence of an $\alpha$-cluster state at high excitation energies in $^{56}$Ni. We used the ideal $\alpha$-gas model and the distributed $m\alpha$-decay ensembles to interpret the experimental findings. The results from both methods, albeit with large error bars, are compatible and can well reproduce the experimental data. The first method places the excitation energy of the $\alpha$ condensate around 96~MeV, whereas the second method yields an excitation energy of $113^{+15}_{-17}$~MeV. As suggested by the results of this experiment, the $\alpha$-cluster resonance state occurs at high-excitation energy in nuclei. Efforts should be made in the future to understand precisely the nature of the condensate state in $\alpha$-conjugate nuclei at high-excitation energies~\cite{Tohsaki2017}. 

\section{Acknowledgements}
\label{thankyou}
This work was supported by the European Commission within the Seventh Framework Program through IA ENSAR (Contract no. RII3-CT-2010-262010), GSI, Darmstadt, Germany, the United States National Science Foundation (Grants No. PHY-1068192 and PHY-1713857), the Hungarian NKFI Foundation (No. K124810), and the Hirao Taro Foundation of the Konan University Association for Academic Research.

%

\begin{thebibliography}{99}


\bibitem{Hoyle_Astro}
F. Hoyle, Astrophys. J. Suppl. Ser. \textbf{1}, 121 (1954)

\bibitem{Ikeda1968}
K. Ikeda et al., Prog. Theor. Phys. Supp. E \textbf{68}, 464 (1968)

\bibitem{Yamada2004}
T. Yamada, and P. Schuck, Phys. Rev. C \textbf{69}, 024309 (2004) 

\bibitem{Itagaki2008}
N. Itagaki et al., Phys. Rev. C \textbf{77},  037301 (2008) 

\bibitem{Freer_Nature}
M. Freer, Nature \textbf{487},  309 (2012) 

\bibitem{Wheeler}
J.A. Wheeler, Phys. Rev. \textbf{52},  1083 (1937)

\bibitem{Horiuchi}
H. Horiuchi, Prog. Theor. Phys. Suppl. \textbf{62}, 90 (1977)

\bibitem{Girod}
M. Girod, and P. Schuck, Phys. Rev. Lett. \textbf{111}, 132503 (2013)

\bibitem{Ebran}
J.P. Ebran et al., Phys. Rev. C \textbf{89}, 031303(R) (2014) 

\bibitem{Elhatisari}
S. Elhatisari et al., Phys. Rev. Lett. \textbf{119}, 222505 (2017)

\bibitem{Tohsaki2001}
A. Tohsaki et al., Phys. Rev. Lett. \textbf{87}, 192501 (2001)

\bibitem{Arickx}
F. Arickx, J. Broeckhove, and E. Deumens, Nucl. Phys. A \textbf{318}, 269 (1979)

\bibitem{Tohsaki2017}
A. Tohsaki et al., Rev. Mod. Phys. \textbf{89}, 011002 (2017)

\bibitem{Freer2018}
M. Freer et al., Rev. Mod. Phys. \textbf{90}, 035004 (2018)

\bibitem{Itoh2011}
M. Itoh et al., Phys. Rev. C \textbf{84}, 054308 (2011)

\bibitem{Smith}
R. Smith et al., Phys. Rev. Lett. \textbf{119}, 132502 (2017) 

\bibitem{Aquila}
D. Dell’ Aquila et al., Phys. Rev. Lett. \textbf{119}, 132501 (2017)

\bibitem{Chernykh}
M. Chernykh et al., Phys. Rev. Lett. \textbf{98}, 032501 (2007)

\bibitem{Uegaki}
E. Uegaki et al., Prog. Theor. Phys. \textbf{57}, 1262 (1977)

\bibitem{Epelbaum}
E. Epelbaum et al., Phys. Rev. Lett. \textbf{106}, 192501 (2011)

\bibitem{Wakasa2002}
T. Wakasa et al., RCNP annual report, Osaka University, Japan (2002)

\bibitem{Wakasa2007}
T. Wakasa et al., Phys. Lett. B \textbf{653}, 173 (2007)

\bibitem{Funaki}
Y. Funaki et al., Phys. Rev. Lett. \textbf{101}, 082502 (2008)

\bibitem{Borderie2016}
B. Borderie et al., Phys. Lett. B \textbf{755}, 475 (2016)

\bibitem{Beck1}
C. Beck (Ed.), Clusters in Nuclei, vol. 1, Lecture Notes in Physics, \textbf{818}, 2010 and references therein.

\bibitem{Beck2}
C. Beck (Ed.), Clusters in Nuclei, vol. 2, Lecture Notes in Physics, \textbf{848}, 2012 and references therein.

\bibitem{Beck3}
C. Beck (Ed.), Clusters in Nuclei, vol. 3, Lecture Notes in Physics, \textbf{875}, 2014 and references therein. 

\bibitem{Itoh_NPA}
M. Itoh et al., Nucl. Phys. A \textbf{738}, 268 (2004)

\bibitem{Freer2012}
M. Freer et al., Phys. Rev. C \textbf{86}, 034320 (2012)

\bibitem{Furuno}
T. Furuno et al., Jour. Phys.: Conf. Ser. \textbf{863}, 012076 (2017)


\bibitem{Akimune2013}
H. Akimune et al., Jour. Phys.: Conf. Ser. \textbf{436}, 012010 (2013)

\bibitem{NNDC}
NNDC(BNL)-Brookhaven National Laboratory, \url{https://www.nndc.bnl.gov}

\bibitem{Anne1987}
R. Anne et al., Nucl. Instrum. Meth. Phys. Res. A \textbf{257}, 215 (1987)

\bibitem{Bagchithesis}
S. Bagchi, Ph.D. thesis, University of Groningen (2015),
\url{https://www.rug.nl/research/portal/files/17683086/Complete_dissertation.pdf}

\bibitem{Demonchy2007}
C. E. Demonchy et al., Nucl. Instrum. Meth. Phys. Res. A \textbf{583}, 341 (2007)

\bibitem{Pancin2012}
J. Pancin et al., Jour. Instrum. \textbf{7}, P01006 (2012) 

\bibitem{Parlog2002}
M. Parlog et al., Nucl. Instrum. Meth. Phys. Res. A \textbf{482}, 693 (2002)


\bibitem{Bagchi2015}
S. Bagchi et al., Phys. Lett. B \textbf{751}, 371 (2015)

\bibitem{Roger2011}
T. Roger et al., Nucl. Instrum. Meth. Phys. Res. A \textbf{638}, 134 (2011) 

\bibitem{Puhlhofer1977}
F. P\"uhlhofer, Nucl. Phys. A \textbf{280}, 267 (1977)

\bibitem{Harakeh1986}
M. N. Harakeh et al., Phys. Lett. B \textbf{176}, 297 (1986)







\end{thebibliography}
%

\end{document}